\documentclass{pasj00}
\draft

\begin{document}
\SetRunningHead{Sato et al.}{Planetary Companions to Evolved
Intermediate-Mass Stars}
\Received{}
\Accepted{}

\title{Planetary Companions to Evolved Intermediate-Mass Stars:
14 Andromedae, 81 Ceti, 6 Lyncis, and HD 167042}



%
 \author{
   Bun'ei \textsc{Sato},\altaffilmark{1}
   Eri \textsc{Toyota},\altaffilmark{2}
   Masashi \textsc{Omiya},\altaffilmark{3}
   Hideyuki \textsc{Izumiura},\altaffilmark{4,5}
   Eiji \textsc{Kambe},\altaffilmark{4}
   Seiji \textsc{Masuda},\altaffilmark{6}
   Yoichi \textsc{Takeda},\altaffilmark{5,7}
   Yoichi \textsc{Itoh},\altaffilmark{2}
   Hiroyasu \textsc{Ando},\altaffilmark{5,7}
   Michitoshi \textsc{Yoshida},\altaffilmark{4,5}
   Eiichiro \textsc{Kokubo},\altaffilmark{5,7}
   and 
   Shigeru \textsc{Ida}\altaffilmark{8}}
 \altaffiltext{1}{Global Edge Institute, Tokyo Institute of Technology,
2-12-1-S6-6 Ookayama, Meguro-ku, Tokyo 152-8550}
 \email{sato.b.aa@m.titech.ac.jp}
 \altaffiltext{2}{Graduate School of Science, Kobe University,
1-1 Rokkodai, Nada, Kobe 657-8501}
 \altaffiltext{3}{Department of Physics, Tokai University,
1117 Kitakaname, Hiratsuka, Kanagawa 259-1292}
 \altaffiltext{4}{Okayama Astrophysical Observatory, National
Astronomical Observatory of Japan, Kamogata,
Okayama 719-0232}
 \altaffiltext{5}{The Graduate University for Advanced Studies,
Shonan Village, Hayama, Kanagawa 240-0193}
 \altaffiltext{6}{Tokushima Science Museum, Asutamu Land Tokushima,
45-22 Kibigadani, Nato, Itano-cho, Itano-gun,
Tokushima 779-0111}
 \altaffiltext{7}{National Astronomical Observatory of Japan, 2-21-1 Osawa,
Mitaka, Tokyo 181-8588}
 \altaffiltext{8}{Department of Earth and Planetary Sciences,
Tokyo Institute of Technology, 2-12-1 Ookayama, Meguro-ku,
Tokyo 152-8551}


\maketitle

\begin{abstract}
We report on the detection of four extrasolar planets orbiting
evolved intermediate-mass stars from a precise Doppler survey
of G--K giants at Okayama Astrophysical Observatory.
All of the host stars are considered to be formerly
early F-type or A-type dwarfs when they were on the main sequence.
14 And (K0 III) is a clump giant with a mass of 2.2 $M_{\odot}$
and has a planet of minimum mass $m_2\sin i=4.8 M_{\rm J}$ in a
nearly circular orbit with a 186 day period. This is one of the
innermost planets around evolved intermediate-mass stars and
such planets have only been discovered in clump giants.
81 Cet (G5 III) is a clump giant with 2.4 $M_{\odot}$
hosting a planet of $m_2\sin i=5.3 M_{\rm J}$ in a 953 day orbit
with an eccentricity of $e=0.21$.
6 Lyn (K0 IV) is a less evolved subgiant with 1.7 $M_{\odot}$
and has a planet of $m_2\sin i=2.4 M_{\rm J}$ in a 899 day orbit
with $e=0.13$.
HD 167042 (K1 IV) is also a less evolved star with 1.5 $M_{\odot}$
hosting a planet of $m_2\sin i=1.6 M_{\rm J}$ in a 418 day orbit
with $e=0.10$. This planet was independently announced by
Johnson et al. (2008, \apj, 675, 784).
All of the host stars have solar or sub-solar
metallicity, which supports the lack of metal-rich tendency in
planet-harboring giants in contrast to the case of dwarfs.
\end{abstract}

\section{Introduction}
Precise Doppler surveys of evolved stars have opened a new
frontier in extrasolar planet searches during the past several
years. Since the successive discoveries of planets around
K-type giants, $\iota$ Dra (Frink et al. 2002) and HD 47536
(Setiawan et al. 2003), and a G-type giant, HD 104985
(Sato et al. 2003), about 20 substellar companions orbiting
evolved stars have been identified so far (Setiawan 2003;
Setiawan et al. 2005; Sato et al. 2007, 2008;
Hatzes et al. 2003, 2005, 2006; Reffert et al. 2006;
Johnson et al. 2007, 2008;
Lovis \& Mayor 2007; Niedzielski et al. 2007;
D$\ddot{\rm{o}}$llinger et al. 2007; Liu et al. 2008).
Planets in evolved stars is now one of the major
subjects in the field of extrasolar planets.

In the past, planets around evolved stars were primarily
studied from the interests in the fate of our solar system,
that is, whether the Earth and the other planets will be
engulfed by the Sun in the future (Sackmann et al. 1993;
Duncan and Lissauer 1998). On the other hand, the current
Doppler surveys of evolved stars have been mainly carried
out in the context of planet searches around intermediate-mass
(1.5--5 $M_{\odot}$) stars. Intermediate-mass stars on the
main sequence, that is early-type dwarfs (B--A dwarfs), are
more difficult for Doppler planet searches because they have
fewer absorption lines in their spectra than later type ones,
which are often broadened due to their rapid rotation, and thus
it is more difficult to achieve a high measurement precision in
radial velocity that is enough to detect orbiting planets
(but see eg. Galland et al. 2005, 2006, which have developed
a technique to extract Doppler information from spectra
of A dwarfs sufficient for detection of substellar companions).
This is actually one of the major reasons
why targets for planet searches had been limited to low-mass
($<1.5M_{\odot}$) F--M dwarfs. On the contrary, late G to
early K giants and subgiants, which are intermediate-mass
stars in evolved stages, have many sharp absorption lines in their
spectra appropriate for precise radial velocity measurements
and their surface activity such as pulsation is quiet enough
not to prevent us from detecting planets. Therefore, these
types of stars have been identified as promising targets
for Doppler planet searches around intermediate-mass stars.

The ongoing planet searches have already revealed that the
properties of planets around evolved intermediate-mass stars
are probably different from those around low-mass dwarfs
(see Bulter et al. 2006 or Udry and Santos 2007, and
references therein for properties of planets around
low-mass stars):
1) high frequency of massive planets (Lovis \& Mayor 2007;
Johnson et al. 2007), which may be supported by the
detection of a planet in an intermediate-mass giant
in the Hyades open cluster (Sato et al. 2007) despite the
absence of planets in low-mass hundred dwarfs in the cluster
(Paulson et al. 2004), 2) lack of inner planets with
orbital semimajor axes of $\lesssim0.7$ AU (Johnson et al.
2007; Sato et al. 2008), and 3) lack of metal-rich
tendency in the host stars of planets (Pasquini et al. 2007;
Takeda et al. 2008).
Although the properties must reflect history of formation
and evolution of planetary systems dependent on their host
stars' mass, they are still in need of confirmation by
a larger number of samples.
Combined with these observational progresses,
planet formation theories applicable to more massive
stars than the Sun have begun to develop rapidly
(Ida and Lin 2005; Burkert and Ida 2007;
Kennedy and Kenyon 2007) as extension of the standard
theory for solar-mass ones (e.g., Hayashi et al. 1985;
Ida and Lin 2004) for the first time since Nakano (1988)
explored the idea nearly 20 years ago.

In this paper, we report on the detection of four planets
orbiting intermediate-mass G--K giants and subgiants
(14 And, 81 Cet, 6 Lyn, and HD 167042) from the Okayama Planet
Search Program (Sato et al. 2005), one of which (HD 167042)
was independently announced by Johnson et al. (2008, \apj, 675,
784). We describe our observations
in Section 2. Properties of the host stars, radial velocity
data, and orbital parameters of their planets are presented
in Section 3. Section 4 is devoted to an analysis of spectral
line shape for the host stars. We summarize our results in
Section 5 with a discussion about the properties of planets
and host stars.

\section{Observations}

Since 2001, we have been conducting a precise Doppler survey of
about 300 G--K giants (Sato et al. 2005) using a 1.88 m telescope,
the HIgh Dispersion Echelle Spectrograph (HIDES; Izumiura 1999),
and an iodine absorption cell (I$_2$ cell; Kambe et al. 2002)
at Okayama Astrophysical Observatory (OAO).
In 2007 December, HIDES was upgraded from single CCD
(2K$\times$4K) to a mosaic of three CCDs, which can simultaneously
cover a wavelength range of 3750--7500${\rm \AA}$ using a RED
cross-disperser (Izumiura et al. in preparation).
For precise radial velocity measurements, we use a wavelength range
of 5000--5800${\rm \AA}$ (covered by the middle CCD after the
upgrade in 2007 December), in which many deep and sharp I$_2$
lines exist. A slit width is set to 200 $\mu$m ($0.76^{\prime\prime}$)
giving a spectral resolution ($R=\lambda/\Delta\lambda$) of 67000
by about 3.3 pixels sampling. We can typically obtain a
signal-to-noise ratio S/N$>$200 pix$^{-1}$ for a $V<6.5$
star with an exposure time shorter than 30 min.
We have achieved a long-term Doppler precision of about 6 m s$^{-1}$
over a time span of 7 years using our own analysis software for
modeling an I$_2$-superposed stellar spectrum (Sato et al. 2002,
2005). Recently, we have succeeded in attaining a Doppler precision
of about 2 m s$^{-1}$ in week-long time scale by improving the
analysis software (Kambe et al. 2008). We are now trying to
maintain the same precision in year-long time scale.

For abundance analysis, we take a pure (I$_2$-free) stellar
spectrum with the same wavelength range and spectral resolution as
those for radial velocity measurements.
We also take a spectrum covering Ca {\small II} H K lines in order
to check the chromospheric activity for stars showing large radial
velocity variations. Although we can take the spectrum simultaneously
with radial velocity data after the upgrade of HIDES, the spectra
presented in this paper except for HD 167042 were obtained before
the upgrade.
In that case, we set the wavelength range to 3800--4500 ${\rm \AA}$
using a BLUE cross-disperser and the slit width to 250 $\mu$m
giving a wavelength resolution of 50000. We typically obtained
S/N $\simeq$ 20 pix$^{-1}$ at the Ca {\small II}
H K line cores for a $B=6$ star with a 30 min exposure.

The reduction of echelle data (i.e. bias subtraction, flat-fielding,
scattered-light subtraction, and spectrum extraction) is performed
using the IRAF\footnote{IRAF is distributed by the National
Optical Astronomy Observatories, which is operated by the
Association of Universities for Research in Astronomy, Inc. under
cooperative agreement with the National Science Foundation,
USA.} software package in the standard way.

\section{Stellar Properties, Radial Velocities, and
Orbital Solutions}

\subsection{14 Andromedae}

\begin{figure}
  \begin{center}
    \FigureFile(85mm,80mm){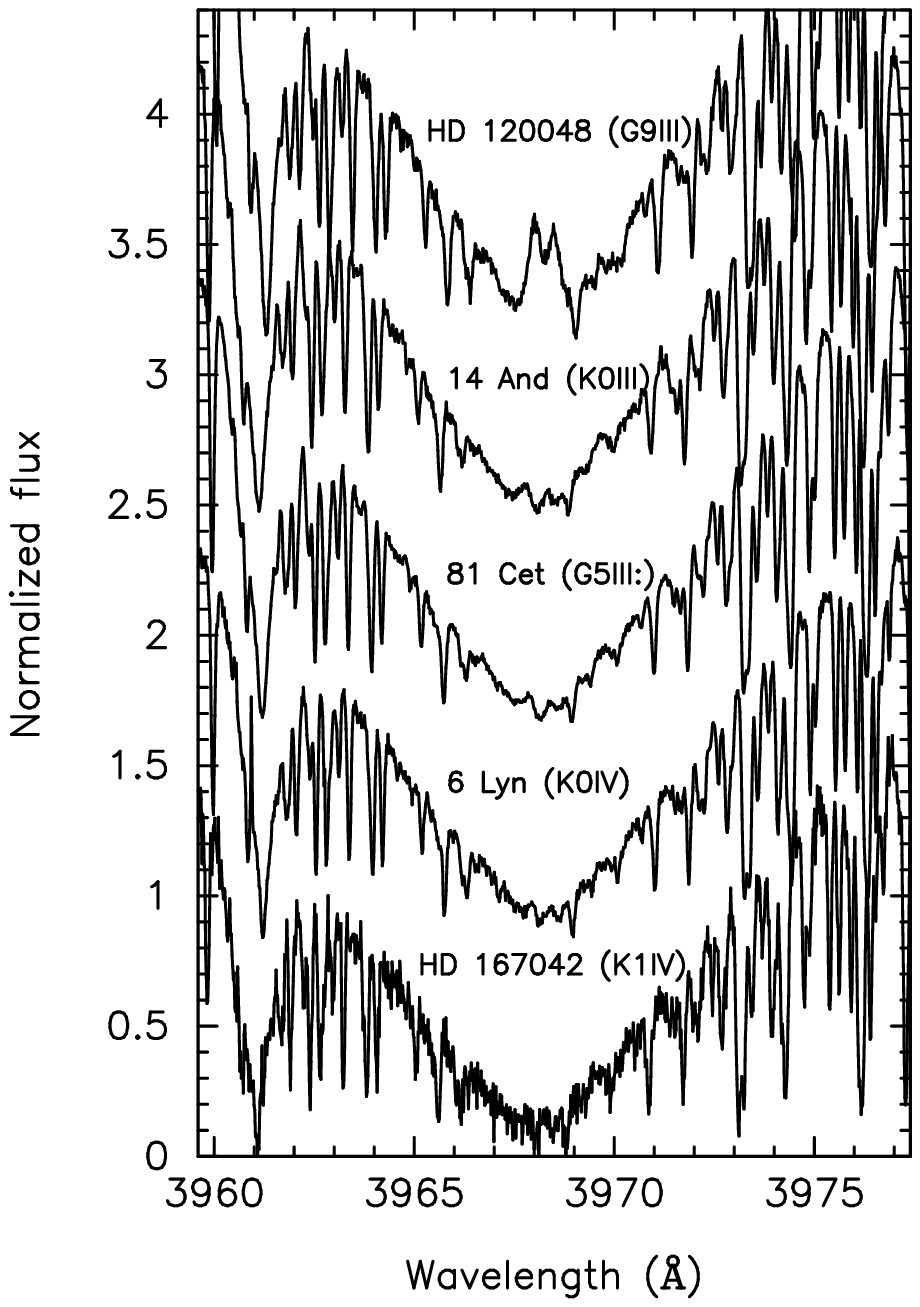}
  \end{center}
\caption{Spectra in the region of Ca H lines. All of the
planet-harboring stars show no significant emissions
in line cores compared to that in the chromospheric active
star HD 120048, which exhibits velocity scatter of about 30 m s$^{-1}$.
A vertical offset of about 0.7 is added to each spectrum.}\label{fig-CaH}
\end{figure}

14 And (HR 8930, HD 221345, HIP 116076) is listed in the Hipparcos
catalog (ESA 1997) as a K0 III giant star with
a $V$ magnitude $V=5.22$ and a color index $B-V=1.029$.
The Hipparcos parallax $\pi=13.09\pm0.71$ mas corresponds
to a distance of 76.4$\pm$4.1 pc and an absolute magnitude $M_V=0.67$
taking account of correction of interstellar extinction $A_V=0.13$
based on the Arenou et al's (1992) table.
{\it Hipparcos} made a total of 80 observations of the star,
revealing a photometric stability down to $\sigma=0.006$ mag.
Ca {\small II} H K lines of the star show no significant
emission in the line cores as shown in the Figure \ref{fig-CaH},
suggesting that the star is chromospherically inactive.

\begin{figure}
  \begin{center}
    \FigureFile(85mm,80mm){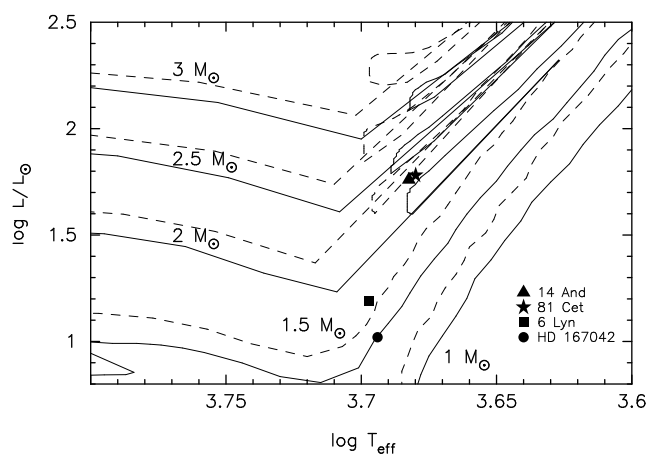}
  \end{center}
\caption{HR diagram of the planet-harboring stars presented in this paper.
Pairs of evolutionary tracks from Lejeune and Schaerer (2001)
for stars with $Z=0.02$ (solar metallicity; solid
lines) and $Z=0.008$ (dashed lines) of masses between 1 and 3
$M_{\odot}$ are also shown.}\label{fig-HRD}
\end{figure}

We derived atmospheric parameters and Fe abundance
of $\sim$ 320 G--K giants including all targets for the Okayama Planet
Search Program based on the spectroscopic approach using the equivalent 
widths of well-behaved Fe I and Fe II lines (cf. Takeda et al. 
2002 for a detailed description of this method; Takeda et al. 2008).
For 14 And, 
$T_{\rm eff}$ = 4813 K,
$\log g$ = 2.63 cm~s$^{-2}$, 
$v_{\rm t}$ = 1.43 km~s$^{-1}$, 
and [Fe/H] = $-$0.24 were obtained.
The bolometric correction was estimated
to $B.C.=-0.33$ based on the Kurucz (1993)'s theoretical calculation.
With use of these parameters and theoretical evolutionary tracks
of Lejeune \& Schaerer (2001), we obtained the
fundamental stellar parameters, $L=58L_{\odot}$, $R=11R_{\odot}$,
and $M=2.2M_{\odot}$, as summarized in Table \ref{tbl-stars}.
The procedure described here is the same as that adopted in
Takeda et al. (2005) (see subsection 3.2 of Takeda et al. (2005)
and Note of Table \ref{tbl-stars} for uncertainties involved
in the stellar parameters).
The projected rotational velocity $v\sin i=2.60$ km s$^{-1}$
was also obtained by Takeda et al. (2008).
Mishenina et al. (2006) obtained
$T_{\rm eff}$ = 4664 K (from line-depth-ratio analysis),
$\log g$ = 2.20 cm~s$^{-2}$, 
$v_{\rm t}$ = 1.4 km~s$^{-1}$, 
[Fe/H] = $-$0.37,
and $M=1.5M_{\odot}$ for the star.
The temperature and [Fe/H] are by $\sim150$ K and by $\sim0.1$ dex
lower than our estimates, respectively. Such differences can
produce $\sim0.5M_{\odot}$ difference in mass of a clump giant
especially for metal-poor case depending on evolutionary models
(see Note of Table 1).
As shown in Figure \ref{fig-HRD}, the star is located at the
clump region on the HR diagram.

\begin{figure}
  \begin{center}
    \FigureFile(85mm,80mm){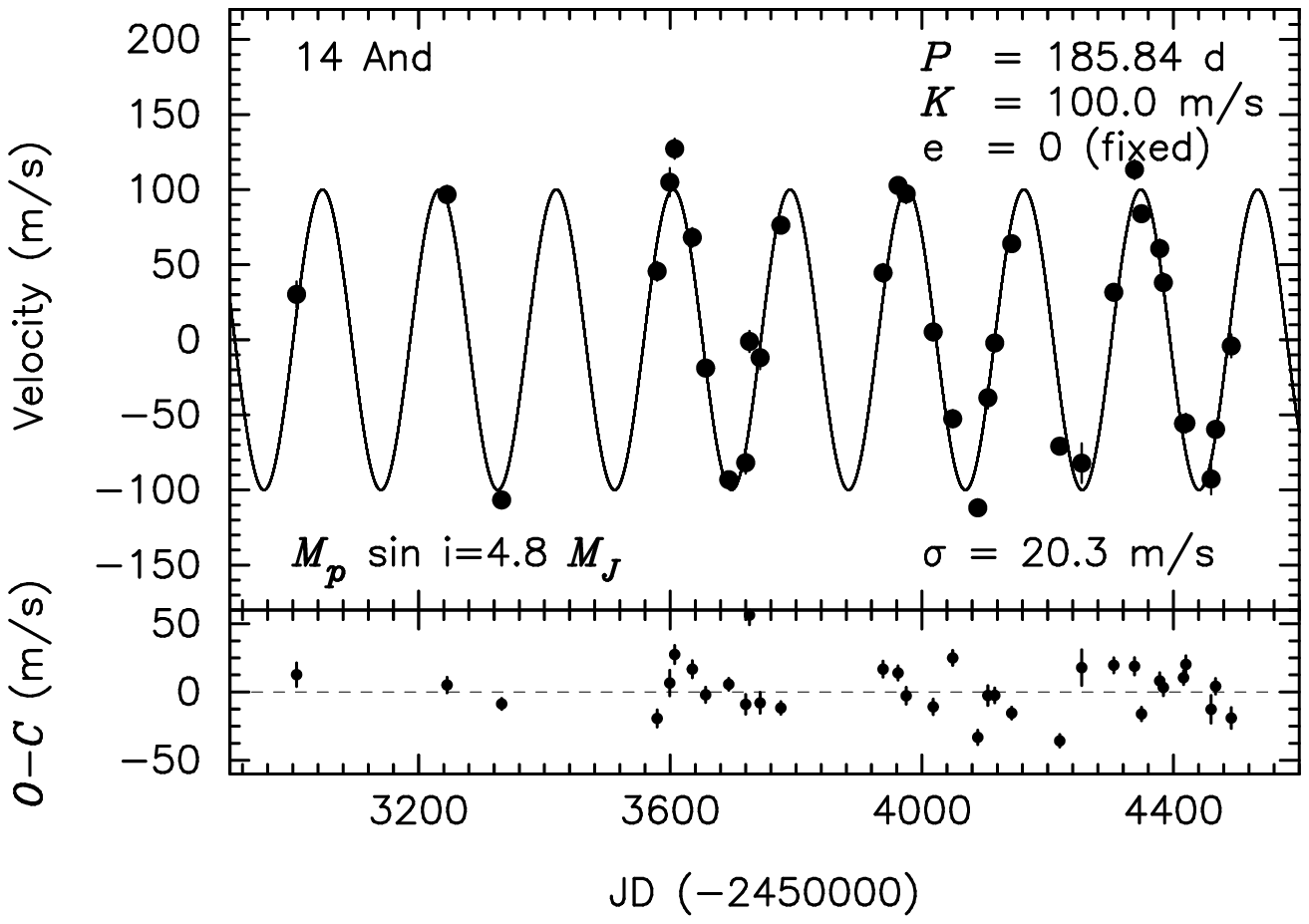}
  \end{center}
\caption{{\it Top}: Observed radial velocities of 14 And (dots). 
The Keplerian orbital fit is shown by the solid line.
{\it Bottom}: Residuals to the Keplerian fit.
The rms to the fit is 20.3 m s$^{-1}$.}
\label{fig-HD221345}
\end{figure}

We collected a total of 34 radial velocity data of 14 And
between 2004 January and 2008 January, with a typical S/N of 200
pix$^{-1}$ for an exposure time of about 600 s.
The observed radial velocities are shown in Figure \ref{fig-HD221345}
and are listed in Table \ref{tbl-HD221345} together with their
estimated uncertainties, which were derived from an ensemble of
velocities from each of $\sim$200 spectral regions
(each 4--5${\rm \AA}$ long) in every exposure. Lomb-Scargle periodogram
(Scargle 1982) of the data exhibits a dominant peak at a period of 188 days.
To assess the significance of this periodicity, we estimated False Alarm
Probability ($FAP$), using a bootstrap randomization method in which the
observed radial velocities were randomly redistributed, keeping fixed
the observation time. We generated 10$^5$ fake datasets in this way,
and applied the same periodogram analysis to them. Since no fake datasets
exhibited a periodogram power higher than the observed dataset,
the $FAP$ is less than $1\times10^{-5}$.

The observed radial velocities can be well fitted by a circular
orbit with a period $P=185.84\pm0.23$ days and a velocity semiamplitude
$K_1=100.0\pm1.3$ m s$^{-1}$.
The resulting model is shown in Figure \ref{fig-HD221345} overplotted
on the velocities, and its parameters are listed in
Table \ref{tbl-planets}.
The uncertainty of each parameter was estimated using a Monte
Carlo approach. We generated 300 fake datasets
by adding random Gaussian noise corresponding to velocity measurement
errors to the observed radial velocities in each set, then found the
best-fit Keplerian parameters for each, and examined the distribution of
each of the parameters.
The rms scatter of the residuals to the Keplerian fit was
20.3 m s$^{-1}$, which is comparable to the scatters of
giants with the same $B-V$ as 14 And in our sample
(Sato et al. 2005).
Adopting a stellar mass of 2.2 $M_{\odot}$, we obtain a minimum mass
for the companion of $m_2\sin i=4.8M_{\rm J}$ and a semimajor axis of
$a=0.83$ AU. The planet is one of the innermost planets ever
discovered around evolved stars.

\subsection{81 Ceti}

81 Cet (HR 771, HD 16400, HIP 12247) is listed in the Hipparcos
catalog (ESA 1997) as a G5 III: giant star with
a $V$ magnitude $V=5.65$, a color index $B-V=1.021$,
and a parallax $\pi=10.29\pm0.97$ mas, corresponding
to a distance of 97.2$\pm$9.2 pc and an absolute magnitude $M_V=0.63$
corrected by interstellar extinction $A_V=0.08$ (Arenou et al. 1992).
{\it Hipparcos} made a total of 58 observations of the star,
revealing a photometric stability down to $\sigma=0.006$ mag.
Ca {\small II} H K lines of the star show no significant
emission in the line cores as shown in the Figure \ref{fig-CaH},
suggesting that the star is chromospherically inactive.
We derived fundamental stellar parameters for the star of
$T_{\rm eff}$ = 4785 K,
$\log g$ = 2.35 cm~s$^{-2}$, 
$v_{\rm t}$ = 1.33 km~s$^{-1}$,
[Fe/H] = $-$0.06,
$L=60L_{\odot}$,
$R=11R_{\odot}$,
and $M=2.4M_{\odot}$, as summarized in Table \ref{tbl-stars}.
As shown in Figure \ref{fig-HRD}, the star is located at the
clump region on the HR diagram.
Mishenina et al. (2006) derived
$T_{\rm eff}$ = 4840 K (from line-depth-ratio analysis),
$\log g$ = 2.50 cm~s$^{-2}$, 
$v_{\rm t}$ = 1.35 km~s$^{-1}$, 
[Fe/H] = $-$0.01,
and $M=2.5M_{\odot}$ for the star, which reasonably agree with
our estimates.

\begin{figure}
  \begin{center}
    \FigureFile(85mm,80mm){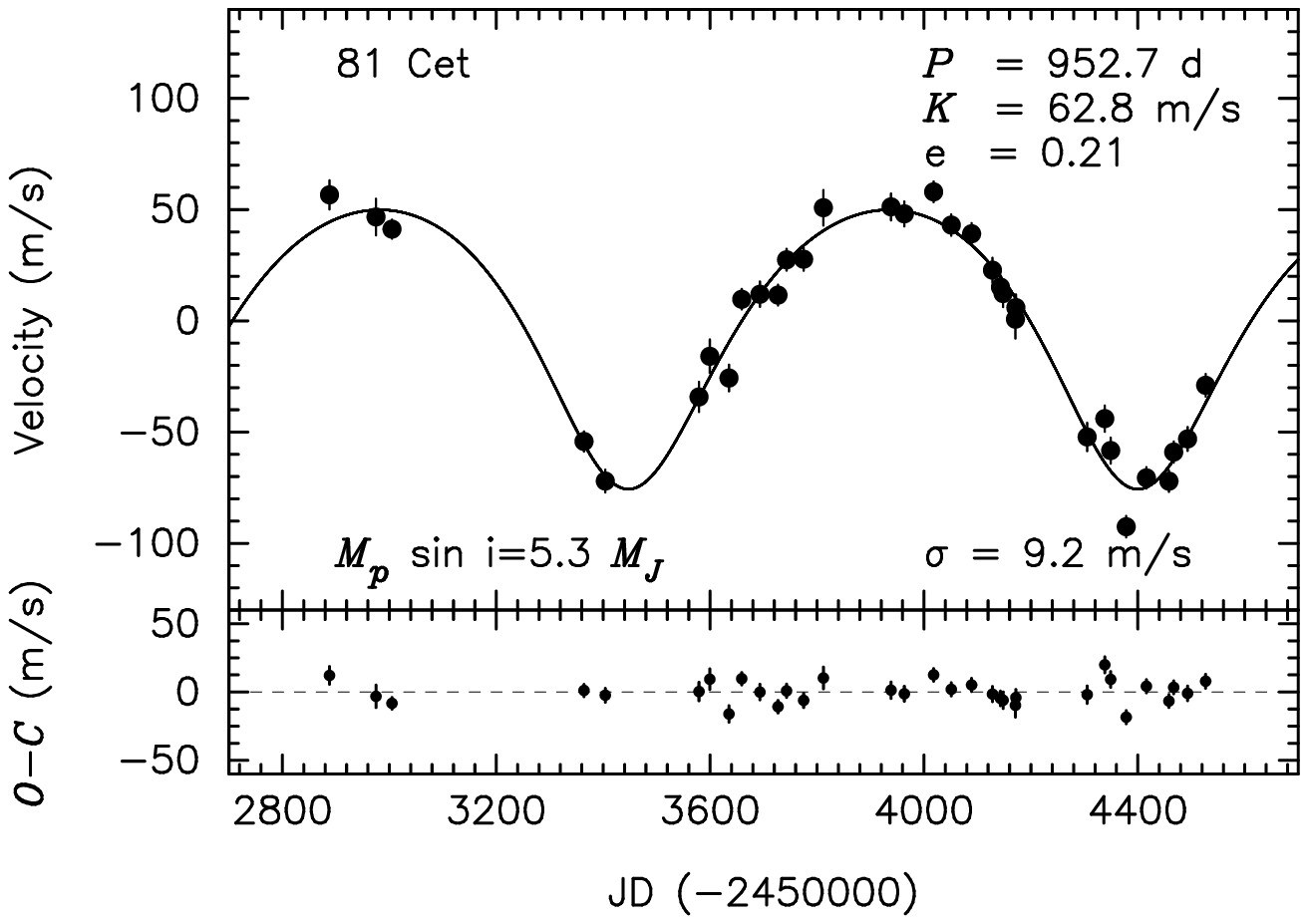}
  \end{center}
\caption{{\it Top}: Observed radial velocities of 81 Cet (dots). 
The Keplerian orbital fit is shown by the solid line.
{\it Bottom}: Residuals to the Keplerian fit.
The rms to the fit is 9.2 m s$^{-1}$.}
\label{fig-HD16400}
\end{figure}

We collected a total of 33 radial velocity data of 81 Cet
between 2003 September and 2008 March, with a typical S/N of 200 pix$^{-1}$
for an exposure time of 900 s.
The observed radial velocities are shown in Figure \ref{fig-HD16400}
and are listed in Table \ref{tbl-HD16400} together with their
estimated uncertainties. Lomb-Scargle periodogram
(Scargle 1982) of the data exhibits a dominant peak at a period of 983 days
with a $FAP<1\times10^{-5}$, which is estimated by the
same way as described in Section 3.1.

The observed radial velocities can be well fitted by a Keplerian orbit
with a period $P=952.7\pm8.8$ days, a velocity semiamplitude
$K_1=62.8\pm1.5$ m s$^{-1}$, and an eccentricity $e=0.206\pm0.029$.
The resulting model is shown in Figure \ref{fig-HD16400} overplotted
on the velocities, and its parameters are listed in
Table \ref{tbl-planets}. The uncertainty of
each parameter was estimated using a Monte Carlo approach as
described in Section 3.1.
The rms scatter of the residuals to the Keplerian fit is
9.2 m s$^{-1}$, which is small compared to the typical scatters of
giants with the same $B-V$ as 81 Cet in our sample
(Sato et al. 2005). Adopting a stellar mass of 2.4 $M_{\odot}$,
we obtain a minimum mass for the companion $m_2\sin i=5.3M_{\rm J}$
and a semimajor axis $a=2.5$ AU. The planet is one of the outermost
planets ever discovered around evolved stars.

\subsection{6 Lyncis}

6 Lyn (HR 2331, HD 45410, HIP 31039) is a less evolved K0
subgiant star with a $V$ magnitude $V=5.86$, a color
index $B-V=0.934$, and the Hipparcos parallax $\pi=17.56\pm0.76$
mas (ESA 1997), placing the star at a distance of 56.9$\pm$2.5 pc.
The distance and an estimated interstellar extinction $A_V=0.03$
(Arenou et al. 1992) yield an absolute magnitude for the star
$M_V=2.05$.
{\it Hipparcos} made a total of 73 observations of the star, revealing
a photometric stability down to $\sigma=0.005$ mag.
Ca {\small II} H K lines of the star show no significant
emission in the line cores as shown in the Figure \ref{fig-CaH},
suggesting that the star is chromospherically inactive.
We derived fundamental stellar parameters for the star of
$T_{\rm eff}$ = 4978 K,
$\log g$ = 3.16 cm~s$^{-2}$, 
$v_{\rm t}$ = 1.10 km~s$^{-1}$, 
[Fe/H] = $-$0.13,
$L=15L_{\odot}$,
$R=5.2R_{\odot}$,
and $M=1.7M_{\odot}$, as summarized in Table \ref{tbl-stars}.
The position of the star on the HR diagram is plotted
in Figure \ref{fig-HRD} based on these parameters.

\begin{figure}
  \begin{center}
    \FigureFile(85mm,80mm){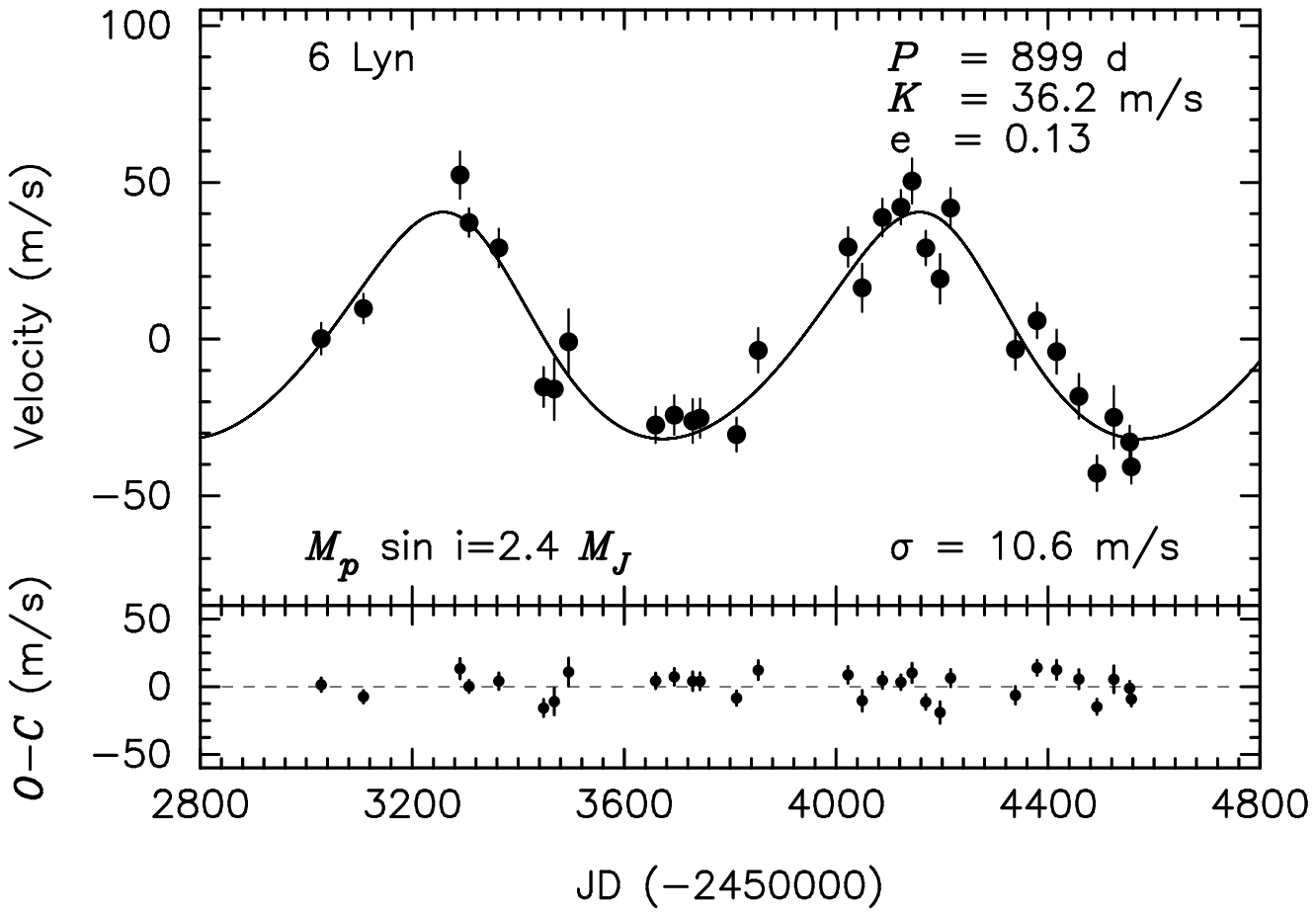}
  \end{center}
\caption{{\it Top}: Observed radial velocities of 6 Lyn (dots). 
The Keplerian orbital fit is shown by the solid line.
{\it Bottom}: Residuals to the Keplerian fit.
The rms to the fit is 10.6 m s$^{-1}$.}
\label{fig-HD45410}
\end{figure}

We collected a total of 30 radial velocity data of 6 Lyn
between 2004 January and 2008 March, with a typical S/N of 200
pix$^{-1}$ for an exposure time of about 1200 s.
The observed radial velocities are shown in Figure \ref{fig-HD45410}
and are listed in Table \ref{tbl-HD45410} together with their
estimated uncertainties. Lomb-Scargle periodogram (Scargle 1982)
of the data exhibits a dominant peak at a period of 917 days
with a $FAP<1\times10^{-5}$, which is estimated by the
same way as described in Section 3.1.

The observed radial velocities can be well fitted by a Keplerian
orbit with a period $P=899\pm19$ days, a velocity semiamplitude
$K_1=36.2\pm1.7$ m s$^{-1}$, and an eccentricity $e=0.134\pm0.052$.
The resulting model is shown in Figure \ref{fig-HD45410}, and its
parameters are listed in Table \ref{tbl-planets}. The uncertainty of
each parameter was estimated using a Monte Carlo approach as
described in Section 3.1.
The rms scatter of the residuals to the Keplerian fit was
10.6 m s$^{-1}$, which is comparable to those for subgiants
(Johnson et al. 2007).
Adopting a stellar mass of 1.7 $M_{\odot}$,
we obtain a minimum mass for the companion of
$m_2\sin i=2.4M_{\rm J}$ and a semimajor axis of
$a=2.2$ AU.

\subsection{HD 167042}

HD 167042 (HR 6817, HIP 89047) is classified in the Hipparcos
catalog (ESA 1997) as a K1 III star with
a $V$ magnitude $V=5.97$ and a color index $B-V=0.943$.
The Hipparcos parallax $\pi=20.00\pm0.51$ mas corresponds to a distance
of 50.0$\pm$1.3 pc and yields an absolute magnitude $M_V=2.47$
corrected by interstellar extinction $A_V=0.01$ (Arenou et al. 1992).
{\it Hipparcos} made a total of 110 observations of the star, revealing
a photometric stability down to $\sigma=0.007$ mag.
Ca {\small II} H K lines of the star show no significant
emission in the line cores, suggesting that the star is
chromospherically inactive (Figure \ref{fig-CaH}).
We derived fundamental stellar parameters for the star of
$T_{\rm eff}$ = 4943 K,
$\log g$ = 3.28 cm~s$^{-2}$, 
$v_{\rm t}$ = 1.07 km~s$^{-1}$, 
[Fe/H] = $+$0.00,
$L=10L_{\odot}$,
$R=4.5R_{\odot}$,
and $M=1.5M_{\odot}$, as summarized in Table \ref{tbl-stars}.
The position of the star on the HR diagram is plotted
in Figure \ref{fig-HRD} based on these parameters.
Judged from the position, the star is considered to be a less
evolved subgiant like 6 Lyn rather than a class III giant.
Johnson et al. (2008) obtained $T_{\rm eff}=5010\pm75$ K,
$\log g$ = 3.47$\pm$0.08 cm~s$^{-2}$, [Fe/H] = $+0.05\pm0.06$,
$M=1.64\pm0.13M_{\odot}$, $R=4.30\pm0.07R_{\odot}$,
and $L=10.5\pm0.05L_{\odot}$ for the star, all of which well
agree with our estimates within the errors.

\begin{figure}
  \begin{center}
    \FigureFile(85mm,80mm){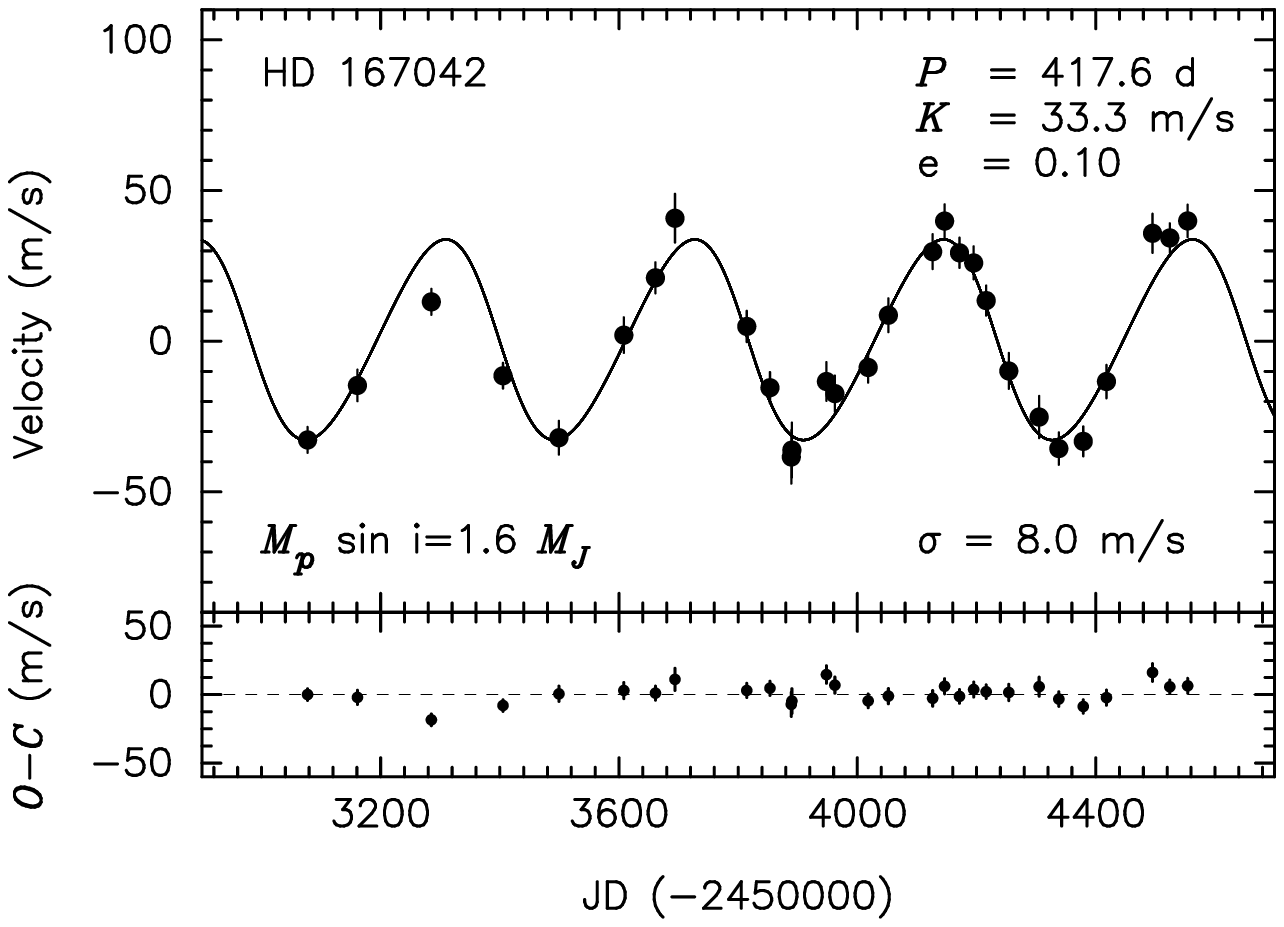}
  \end{center}
\caption{{\it Top}: Observed radial velocities of HD 167042 (dots). 
The Keplerian orbital fit is shown by the solid line.
{\it Bottom}: Residuals to the Keplerian fit.
The rms to the fit is 8.0 m s$^{-1}$.}
\label{fig-HD167042}
\end{figure}

A planetary companion to HD 167042 was recently reported
by Johnson et al. (2008). We collected a total of 29 radial
velocity data of the star between 2004 March and 2008 March,
which is almost the same period of time as that of Johnson et al.
The data have typical S/N of 200 pix$^{-1}$ for exposure time
of about 1500 s.
The observed radial velocities are shown in Figure \ref{fig-HD167042}
and are listed in Table \ref{tbl-HD167042} together with their
estimated uncertainties. Lomb-Scargle periodogram (Scargle 1982)
of the data exhibits a dominant peak at a period of 432 days
with a $FAP<1 \times 10^{-5}$, which is estimated by the
same way as described in Section 3.1.

The observed radial velocities can be well fitted by a Keplerian
orbit with a period $P=417.6\pm4.5$ days, a velocity
semiamplitude $K_1=33.3\pm1.6$ m s$^{-1}$, and an eccentricity
$e=0.101\pm0.066$.
The resulting model is shown in Figure \ref{fig-HD167042}, and its
parameters are listed in Table \ref{tbl-planets}. The uncertainty
of each parameter was estimated using a Monte Carlo approach
as described in Section 3.1.
The rms scatter of the residuals to the Keplerian fit was
8.0 m s$^{-1}$, which is consistent with Johnson et al.
(2008)'s result.
Adopting a stellar mass of 1.5 $M_{\odot}$, we obtain a minimum
mass for the companion of $m_2\sin i=1.6M_{\rm J}$ and a semimajor
axis of $a=1.3$ AU. All of the parameters are in good agreement
with those obtained by Johnson et al. (2008) and thus we
have independently confirmed the existence of the planet.

\section{Line Shape Analysis}

To investigate other causes producing apparent radial velocity
variations such as pulsation and rotational modulation rather
than orbital motion, spectral line shape analysis was performed
with use of high resolution stellar templates followed by the
technique of Sato et al. (2007). In our technique, we extract
a high resolution iodine-free stellar template from several stellar
spectra contaminated by iodine lines (Sato et al. 2002).
Basic procedure of the technique is as follows; first, we model observed
star+I$_2$ spectrum in a standard manner but using the initial
guess of the intrinsic stellar template spectrum.
Next we take the difference between the observed star+I$_2$
spectrum and the modeled one. Since the difference is mainly considered
to be due to an imperfection of the initial guess of the stellar
template spectrum, we revise the initial guess taking account of the
difference and model the observed star+I$_2$ spectrum using the revised
guess of the template. We repeat this process until we obtain sufficient
agreement between observed and modeled spectrum. We take average
of thus obtained stellar templates from several observed star+I$_2$
spectra to increase S/N ratio of the template. Details of this
technique are described in Sato et al. (2002).

For spectral line shape analysis, we extracted two stellar templates
from 5 star+I$_2$ spectra at the peak and valley phases of observed
radial velocities for each star. Then, cross correlation
profiles of the two templates were calculated for 50--80 spectral
segments (4--5${\rm \AA}$ width each) in which severely blended
lines or broad lines were not included.
Three bisector quantities were calculated for the cross correlation
profile of each segment: the velocity span (BVS), which is the
velocity difference between two flux levels of the bisector;
the velocity curvature (BVC), which is the difference of the
velocity span of the upper half and lower half of the bisector;
and the velocity displacement (BVD), which is the average of
the bisector at three different flux levels.
We used flux levels of 25\%, 50\%, and 75\% of the cross
correlation profile to calculate the above quantities.
Resulting bisector quantities for each star are listed in
Table \ref{tbl-bisector}.
As expected from the planetary hypothesis, both of the BVS and
the BVC for 81 Cet, 6 Lyn, and HD 167042 are identical to zero,
which means that the cross
correlation profiles are symmetric, and the average BVD is
consistent with the velocity difference between the two templates
at the peak and valley phases of observed radial velocities
($\simeq 2K_1$).
The BVS for 14 And is about 20 m s$^{-1}$, which is
large compared to those for other stars,
suggesting the higher intrinsic variability and possible variations
in the line profiles for the star.
This may be consistent with the large rms scatters of
the residuals to the Keplerian fit ($\sigma=20.3$ m s$^{-1}$)
and higher rotational velocity ($v\sin i=2.6$ km s$^{-1}$)
for the star. However, the BVD value ($-$177 m s$^{-1}$) is
consistent with the velocity difference between the two
templates ($\simeq 2K_1$) and the BVS value is only about
one ninth of the BVD. Thus it is unlikely that the observed
radial velocity variations are produced by changes in the
line shape due to intrinsic activity such as pulsation or
rotational modulation.

Based on these results, we conclude that the radial velocity
variability observed in these 4 stars are best explained by
orbital motion, although line shape variability for 14 And
deserves to be investigated in more detail.

\section{Summary and Discussion}

We here reported four new planets around evolved intermediate-mass
stars from the Okayama Planet Search Program; two of them orbit
clump giants and the other two orbit subgiants. In total, we
discovered 7 planets and 1 brown dwarf around intermediate-mass
clump giants (2.1--2.7$M_{\odot}$) and 2 planets around
subgiants (1.5 and 1.7$M_{\odot}$) so far from the program.
The host stars are considered to be formerly early F-type
or A-type dwarfs when they were on the main sequence.

Like all of the other planets found around these types of
stars, the four planets presented in this paper reside
beyond $\sim0.7$ AU from the central stars.
Since many planets are known to exist within 0.7 AU around
low-mass dwarfs, the lack of inner planets around evolved
intermediate-mass stars must reflect different
history of formation and evolution of planetary systems.
Two scenarios can account for the orbital distribution;
one is that inner planets are primordially deficient
around intermediate-mass stars and the other is that
they have been engulfed by the expanding central stars
due to stellar evolution. Mass loss of the central star
makes planets move outward because of their weakened
gravitational pull on the planets, but orbital change
due to mass loss is negligible in RGB phase for planets
around intermediate-mass stars because the mass loss of
those stars in RGB phase is negligible (see discussion
in Sato et al. 2008).
The lack of inner planets around
less evolved subgiants (Johnson et al. 2007, 2008)
may favor the former scenario.
Since there is no observational bias against detecting
planets with small semimajor axes around subgiants,
whose radii are $\sim5R_{\odot}$ ($=0.025$ AU) at most
and their intrinsic variability in radial velocity is
adequately small ($\lesssim10$ m s$^{-1}$), the lack of
inner planets around them
may be primordial. On the other hand, as for planets
around intermediate-mass clump giants, evolutionary effect
of the central stars can not be ignored.
When we assume that many of clump giants are post-RGB stars
(core-helium-burning stars), planets in orbits with
$\lesssim0.5$ AU could have been engulfed by the central
stars at the tip of RGB ($R_{\star}\sim25$--$40R_{\odot}$ for
a 2--3 $M_{\odot}$ star) due to tidal force from the
central stars (Sato et al. 2008). Thus we can not reject
the possibility that planets had originally existed in
short orbital distances around progenitors of clump giants.
It should be noted that all of the four innermost planets
around evolved intermediate-mass stars with semimajor axes
of 0.7--1 AU have been discovered around clump giants.
Excluding the planets around clump giants, all of the six
planets ever discovered around intermediate-mass subgiants
reside beyond 1 AU, which expands the lack of inner
planets around them. Although the number is still small,
properties of planets might be different between the two
populations when we consider that clump giants include
$\gtrsim2M_{\odot}$ stars while subgiants are limited up to
$\lesssim2M_{\odot}$. Larger number of planets,
which will be provided by ongoing Doppler surveys of evolved
stars among the globe, enables us to derive their statistical
properties more clearly.

All of the host stars of the four planets presented here
have solar or sub-solar metallicity. The results
support the lack of metal-rich tendency in planet-harboring
giants recently claimed by several authors (e.g., Pasquini et al.
2007; Takeda et al. 2008), which makes marked contrast to the
case of dwarf stars, where planet-harboring stars tend to
be generally metal-rich (see e.g., Gonzalez 2003 or Udry
and Santos 2007, and references therein). Interpretation
of the absence of metal-rich trend in planet-harboring
giants still remains to be cleared. High efficiency
in core-accretion in massive proto-planetary disks around
massive stars combined with high efficiency in inward orbital
migration of planets (finally falling into the central stars)
in metal-rich disks might reproduce
the lack of metal-rich trend. Alternatively,
metallicity-independent planet formation scenario such as disk
instability model (e.g., Boss 2002) might be possible.
Pasquini et al. (2007) pointed out a possibility that
metallicity excess in planet-harboring dwarfs originates
from accretion of metal-rich material and the excess
is diluted by deep convective envelope at the stage of giants.
It should be noted, however, that there are no super-metal-rich
stars with [Fe/H]$>+0.2$ in our sample regardless of
the existence of orbiting planets (Takeda et al. 2008). Since the
frequency distribution of planets steeply rises in
[Fe/H]$>+0.2$ for solar-type dwarfs (Fischer \& Valenti 2005),
we might see metal-poor ([Fe/H]$<+0.2$) tail of the
same distribution, where frequency of planets is less sensitive
to metallicity, in the case of giants. Detailed investigation
of metallicity distribution for a population of giants is required to
derive a firm conclusion on planet-metallicity correlation in giants
and its origin.

Host stars' metallicity could control the fate of orbiting planets
as well as their birth.
From the view point of the engulfment scenario,
inner planets could be deficient around metal-rich clump giants
of $\sim2M_{\odot}$ compared to around metal-poor ones.
A metal-rich star with $\sim2M_{\odot}$ can have more than two
times larger maximum radius ($>70R_{\odot}$) than a metal-poor
one at the tip of RGB
(cf. evolutionary tracks by Girardi et al. 2000; Lejeune and
Schaerer 2001; Claret 2004, 2006, 2007), which may increase a
chance for the central star to engulf inner planets even up
to $\sim1$ AU. However, the current knowledge about evolutionary tracks
of giants is not sufficient to estimate this effect precisely.
Stellar evolutionary track at RGB phase for a $\sim2M_{\odot}$
star is sensitive to slight difference in stellar mass and
metallicity because this mass range is a border where
helium burning starts in a degenerated core or a non-degenerated one
and thus the tracks are largely dependent on adopted models.
To investigate orbital distribution of planets around post-RGB stars
together with precise determination of stellar mass (by using technique
of asteroseismology; e.g., Frandsen et al. 2002;
Hatzes \& Zechmeister 2007; Ando et al. 2008) and metallicity would
conversely give constraints on stellar evolution for
this problematic mass range from observational side.

As described above, the recent discoveries of planets around evolved
stars have made stellar evolution and fate of planets a renewed
area of study.
This subject has been studied by several authors (Sackmann et al.
1993; Duncan and Lissauer 1998; Rasio et al. 1996; Siess and Livio
1999ab; Villanver and Livio 2007) mainly focusing on a solar-mass
star based on the interests in the future of the solar system,
and/or an asymptotic giant branch star with the scope of planets
in planetary nebula and white dwarfs.
Siess and Livio (1999b) proposed that the IR excess
and high Li abundance observed in a few percent of G--K giants
originate from the accretion of a substellar companion.
Now is the time to promote such kinds of studies more
extensively not only for solar-mass stars but also for
intermediate-mass ones especially in RGB phase by both
theoretical and observational approaches.
\\

This research is based on data collected at Okayama Astrophysical
Observatory (OAO), which is operated by National Astronomical
Observatory of Japan (NAOJ).
We are grateful to all the staff members of OAO for their support
during the observations. 
Data analysis was in part carried out on ``sb'' computer system
operated by the Astronomy Data Center (ADC) and
Subaru Telescope of NAOJ. We thank the National Institute
of Information and Communications Technology for their support
on high-speed network connection for data transfer and analysis.
B.S. is supported by MEXT's program "Promotion of Environmental
Improvement for Independence of Young Researchers" under the Special
Coordination Funds for Promoting Science and Technology.
H.I. and M.Y. are supported by Grant-in-Aid for
Scientific Research (C) No.13640247 and (B) No.18340055, respectively,
from the Japan Society for the Promotion of Science (JSPS).
This research has made use of the SIMBAD database, operated at
CDS, Strasbourg, France.


\newpage

\onecolumn
\begin{table}[h]
\caption{Stellar parameters}\label{tbl-stars}
\begin{center}
\begin{tabular}{cccccc}\hline\hline
Parameter      & 14 And & 81 Cet & 6 Lyn & HD 167042\\
\hline			   			   
Sp. Type         & K0 III   & G5 III:         & K0 IV             & $^{\dagger}$K1 IV\\
$\pi$ (mas)      & 13.09$\pm$0.71 & 10.29$\pm$0.97  & 17.56$\pm$0.76    & 20.00$\pm$0.51\\
$V$              & 5.22  & 5.65            & 5.86              & 5.97               \\
$B-V$            & 1.029      & 1.021           & 0.934             & 0.943          \\
$A_{V}$          & 0.13    & 0.08            & 0.03              & 0.01             \\
$M_{V}$          & 0.67   & 0.63            & 2.05              & 2.47             \\
$B.C.$           & $-$0.33   & $-$0.34         & $-$0.27           & $-$0.28          \\
$T_{\rm eff}$ (K) & 4813$\pm$20   & 4785$\pm$25     & 4978$\pm$18       & 4943$\pm$12        \\ 
$\log g$         & 2.63$\pm$0.07   & 2.35$\pm$0.08   & 3.16$\pm$0.05     & 3.28$\pm$0.05      \\
$v_t$            & 1.43$\pm$0.09   & 1.33$\pm$0.06   & 1.10$\pm$0.07     & 1.07$\pm$0.07      \\
$[$Fe/H$]$       & $-$0.24$\pm$0.03 & $-$0.06$\pm$0.03 & $-$0.13$\pm$0.02   & $+$0.00$\pm$0.02    \\
$L$ ($L_{\odot}$) & 58   & 60               & 15                 & 10                  \\
$R$ ($R_{\odot}$) & 11 (10--12)   & 11 (10--13)     & 5.2 (4.9--5.6)    & 4.5 (4.3--4.7)     \\
$M$ ($M_{\odot}$) & 2.2 (2.0--2.3) & 2.4 (2.0--2.5)  & 1.7 (1.5--1.8)   & 1.5 (1.3--1.7)   \\
$v\sin i$ (km s$^{-1}$) & 2.60     & 1.80     & 1.32              & 0.67               \\
\hline
\end{tabular}
\end{center}
$^{\dagger}$ The star is listed in the Hipparcos catalogue as a K1 III giant.
But judged from the position of the star on the HR diagram (Figure \ref{fig-HRD}),
the star should be better classified as a less evolved subgiant.
\\
Note -- The uncertainties of [Fe/H], $T_{\rm eff}$, $\log g$,
and $v_{\rm t}$, are internal statistical errors (for a given data set
of Fe~{\sc i} and Fe~{\sc ii} line equivalent widths) evaluated
by the procedure described in subsection 5.2 of Takeda et al. (2002).
Since these parameter values are sensitive to slight changes in the
equivalent widths as well as to the adopted set of lines (Takeda et al.
2008), realistic ambiguities may be by a factor of $\sim$ 2--3 larger
than these estimates from a conservative point of view
(e.g., 50--100 $K$ in $T_{\rm eff}$, 0.1--0.2~dex in $\log g$).
Values in the parenthesis for stellar radius and mass correspond to
the range of the values assuming the realistic uncertainties in
$\Delta\log L$ corresponding to parallax errors in the Hipparcos
catalog, $\Delta\log T_{\rm eff}$ of $\pm0.01$ dex ($\sim\pm100$~K),
and $\Delta{\rm [Fe/H]}$ of $\pm0.1$ dex.
The resulting mass value may also appreciably depend on the chosen
set of theoretical evolutionary tracks (e.g., the systematic
difference as large as $\sim 0.5M_{\odot}$ for the case of
metal-poor tracks between Lejeune \& Schaerer (2001) and
Girardi et al. (2000).; see also footnote 3 in Sato et al. 2008).
\end{table}

\begin{longtable}{ccc}
  \caption{Radial Velocities of 14 And}\label{tbl-HD221345}
  \hline\hline
  JD & Radial Velocity & Uncertainty\\
  ($-$2450000) & (m s$^{-1}$) & (m s$^{-1}$)\\
  \hline
  \endhead
3005.9546 & 30.2 & 8.6\\
3245.2465 & 96.8 & 5.7\\
3332.0661 & $-$106.7 & 4.0\\
3579.1759 & 45.6 & 6.2\\
3599.3078 & 104.9 & 9.3\\
3607.1357 & 127.2 & 6.6\\
3635.2106 & 68.1 & 6.2\\
3656.1741 & $-$18.8 & 5.5\\
3693.0493 & $-$93.2 & 4.3\\
3719.9577 & $-$81.8 & 7.2\\
3726.0131 & $-$1.1 & 7.0\\
3742.9362 & $-$11.9 & 7.6\\
3775.9237 & 76.3 & 4.6\\
3938.2816 & 44.6 & 5.9\\
3962.2592 & 102.7 & 5.2\\
3975.1040 & 97.1 & 6.2\\
4018.0820 & 5.3 & 5.6\\
4049.1049 & $-$52.6 & 5.5\\
4089.0127 & $-$111.9 & 5.2\\
4104.8910 & $-$38.6 & 7.3\\
4115.8872 & $-$2.1 & 5.4\\
4142.9051 & 64.0 & 4.4\\
4219.2924 & $-$70.8 & 4.6\\
4254.2487 & $-$82.1 & 13.1\\
4305.1747 & 31.6 & 5.5\\
4338.1549 & 113.3 & 6.3\\
4349.1366 & 83.9 & 4.9\\
4378.1594 & 60.7 & 5.9\\
4384.0522 & 38.0 & 6.0\\
4415.9392 & $-$55.8 & 5.0\\
4419.9316 & $-$55.5 & 6.3\\
4460.0025 & $-$92.7 & 10.1\\
4466.9496 & $-$59.7 & 5.7\\
4491.8965 & $-$4.1 & 7.6\\
  \hline
\end{longtable}

\begin{table}
  \caption{Orbital Parameters}\label{tbl-planets}
  \begin{center}
    \begin{tabular}{lrrrrrr}
  \hline\hline
  Parameter      &14 And & 81 Cet & 6 Lyn & HD 167042\\
  \hline
$P$ (days)                    & 185.84$\pm$0.23   & 952.7$\pm$8.8     & 899$\pm$19         & 417.6$\pm$4.5\\
$K_1$ (m s$^{-1}$)             & 100.0$\pm$1.3    & 62.8$\pm$1.5      & 36.2$\pm$1.7       & 33.3$\pm$1.6\\
$e$                           & 0 (fixed)         & 0.206$\pm$0.029   & 0.134$\pm$0.052    & 0.101$\pm$0.066\\
$\omega$ (deg)                & 0 (fixed)         & 175.0$\pm$6.9     & 27$\pm$27          & 82$\pm$52\\
$T_p$    (JD$-$2,450,000)     & 2861.4$\pm$1.5    & 2486$\pm$26       & 3309$\pm$60        & 2974$\pm$60\\
$a_1\sin i$ (10$^{-3}$AU)     & 1.712$\pm$0.024    & 5.39$\pm$0.12     & 2.97$\pm$0.16      & 1.274$\pm$0.057\\
$f_1(m)$ (10$^{-6}M_{\odot}$)  & 1.936$\pm$0.079    & 2.30$\pm$0.15     & 0.431$\pm$0.062    & 0.158$\pm$0.022\\
$m_2\sin i$ ($M_{\rm J}$)      & 4.8               & 5.3               & 2.4                & 1.6\\
$a$ (AU)                      & 0.83              & 2.5               & 2.2                & 1.3\\
$N_{\rm obs}$                  & 34                & 33                & 30                 & 29\\
rms (m s$^{-1}$)              & 20.3              & 9.2               & 10.6               & 8.0\\
Reduced $\sqrt{\chi^2}$       & 3.3              & 1.7               & 1.6                & 1.5\\
  \hline
    \end{tabular}
  \end{center}
\end{table}

\begin{longtable}{ccc}
  \caption{Radial Velocities of 81 Cet}\label{tbl-HD16400}
  \hline\hline
  JD & Radial Velocity & Uncertainty\\
  ($-$2450000) & (m s$^{-1}$) & (m s$^{-1}$)\\
  \hline
  \endhead
2888.2622 & 56.6 & 6.5\\
2975.1133 & 46.7 & 8.2\\
3005.0714 & 41.2 & 4.3\\
3364.0728 & $-$54.2 & 4.5\\
3403.9822 & $-$72.0 & 5.1\\
3579.2561 & $-$34.2 & 6.8\\
3599.3232 & $-$15.9 & 7.5\\
3635.2973 & $-$25.7 & 6.0\\
3659.2479 & 9.7 & 4.6\\
3693.1606 & 12.0 & 5.7\\
3727.0642 & 11.6 & 4.7\\
3743.0368 & 27.4 & 4.9\\
3774.9443 & 27.7 & 5.2\\
3811.9192 & 50.8 & 8.0\\
3938.3038 & 51.2 & 6.0\\
3963.2839 & 48.1 & 5.6\\
4018.1630 & 58.0 & 4.6\\
4051.1382 & 43.0 & 4.8\\
4089.0958 & 39.1 & 4.8\\
4127.9761 & 22.8 & 5.5\\
4142.9929 & 15.1 & 5.3\\
4147.9770 & 12.3 & 6.0\\
4170.9343 & 0.7 & 8.7\\
4171.9188 & 6.0 & 5.9\\
4305.2794 & $-$52.2 & 6.4\\
4338.2888 & $-$44.0 & 5.9\\
4349.2596 & $-$58.3 & 5.8\\
4378.1919 & $-$92.5 & 4.8\\
4416.0384 & $-$70.6 & 4.8\\
4458.0891 & $-$72.1 & 4.8\\
4467.0517 & $-$59.0 & 4.7\\
4493.0207 & $-$53.1 & 5.4\\
4526.9147 & $-$28.9 & 5.1\\
  \hline
\end{longtable}

\begin{longtable}{ccc}
  \caption{Radial Velocities of 6 Lyn}\label{tbl-HD45410}
  \hline\hline
  JD & Radial Velocity & Uncertainty\\
  ($-$2450000) & (m s$^{-1}$) & (m s$^{-1}$)\\
  \hline
  \endhead
3028.2002 & 0.2 & 5.0\\
3107.9650 & 9.8 & 4.7\\
3290.2429 & 52.4 & 7.5\\
3307.1619 & 37.2 & 4.5\\
3363.3240 & 29.1 & 6.1\\
3448.0390 & $-$15.3 & 6.3\\
3467.9581 & $-$15.9 & 9.8\\
3494.9817 & $-$0.8 & 10.3\\
3659.2879 & $-$27.4 & 5.7\\
3694.3102 & $-$24.2 & 6.2\\
3729.1980 & $-$26.1 & 6.9\\
3743.1611 & $-$25.2 & 6.2\\
3811.9435 & $-$30.4 & 5.3\\
3852.9669 & $-$3.6 & 7.1\\
4022.3010 & 29.4 & 6.2\\
4049.2443 & 16.4 & 7.7\\
4087.2793 & 38.8 & 6.0\\
4122.2054 & 42.1 & 5.5\\
4143.1678 & 50.4 & 7.2\\
4169.0485 & 29.1 & 5.5\\
4196.0062 & 19.2 & 7.9\\
4215.9689 & 41.8 & 6.3\\
4338.3163 & $-$3.3 & 6.5\\
4379.2591 & 5.9 & 5.6\\
4416.1067 & $-$4.0 & 7.0\\
4458.1686 & $-$18.3 & 7.1\\
4492.1646 & $-$42.8 & 5.6\\
4524.0101 & $-$24.9 & 9.9\\
4553.9338 & $-$32.8 & 5.2\\
4556.9840 & $-$40.8 & 5.3\\
  \hline
\end{longtable}

\begin{longtable}{ccc}
  \caption{Radial Velocities of HD 167042}\label{tbl-HD167042}
  \hline\hline
  JD & Radial Velocity & Uncertainty\\
  ($-$2450000) & (m s$^{-1}$) & (m s$^{-1}$)\\
  \hline
  \endhead
3077.3248 & $-$32.8 & 4.3\\
3161.1350 & $-$14.7 & 5.2\\
3285.0198 & 13.1 & 4.3\\
3405.3778 & $-$11.5 & 4.3\\
3499.1405 & $-$32.0 & 5.6\\
3608.0945 & 2.0 & 5.9\\
3661.0231 & 21.0 & 5.1\\
3693.9118 & 40.8 & 8.1\\
3814.3203 & 4.9 & 5.2\\
3853.2342 & $-$15.4 & 5.2\\
3889.2354 & $-$38.4 & 8.8\\
3890.1815 & $-$36.1 & 9.1\\
3948.1031 & $-$13.4 & 6.4\\
3962.1760 & $-$17.4 & 6.0\\
4018.0037 & $-$8.7 & 5.0\\
4051.9409 & 8.6 & 5.6\\
4126.3613 & 29.7 & 5.8\\
4146.3220 & 39.9 & 5.5\\
4171.2461 & 29.4 & 5.0\\
4195.2715 & 26.0 & 5.5\\
4216.2184 & 13.5 & 5.0\\
4254.1603 & $-$9.9 & 5.9\\
4305.0598 & $-$25.2 & 7.0\\
4338.0740 & $-$35.6 & 5.3\\
4379.0062 & $-$33.2 & 5.0\\
4417.8831 & $-$13.4 & 5.5\\
4495.3581 & 35.8 & 6.5\\
4524.3108 & 34.3 & 4.9\\
4554.3268 & 39.9 & 5.4\\
  \hline
\end{longtable}

\begin{table}
  \caption{Bisector Quantities}\label{tbl-bisector}
  \begin{center}
    \begin{tabular}{lrrrrrr}
  \hline\hline
  Bisector Quantities & 14 And & 81 Cet & 6 Lyn & HD 167042\\
  \hline
Bisector Velocity Span (BVS) (m s$^{-1}$) & 19.8$\pm$5.8 & 2.4$\pm$3.0     & $-$5.5$\pm$5.1    & 1.0$\pm$4.3\\
Bisector Velocity Curve (BVC) (m s$^{-1}$) & $-$1.6$\pm$2.8 & $-$2.8$\pm$2.1     & $-$0.3$\pm$4.1       & 1.9$\pm$3.1\\
Bisector Velocity Displacement (BVD) (m s$^{-1}$) & $-$177.0$\pm$11.2 & $-$115.3$\pm$7.4  & $-$60.9$\pm$8.9  & $-$58.9$\pm$8.0\\
  \hline
    \end{tabular}
  \end{center}
\end{table}

\end{document}